\documentclass[twocolumn,aps,pra,groupedaddress,floatfix,notitlepage,showpacs,nobalancelastpage]{revtex4}
\usepackage{graphicx, amsmath,amssymb,bm,color,engord}
\usepackage[colorlinks,citecolor=red,urlcolor=blue]{hyperref}
\newcommand{\kfup}{k_{\rm F\uparrow}}
\newcommand{\kfdown}{k_{\rm F\downarrow}}

\newcommand{\as}{a^{}_s}

\newcommand{\tc}{{T_{\rm c}}}

\newcommand{\efup}{\epsilon_{{\rm F}\uparrow}}\newcommand{\efdown}{\epsilon_{{\rm F}\downarrow}}

\newcommand{\curG}{{\cal G}}

\newcommand{\bk}{{\bf k}}

\newcommand{\Sigmah}{\hat{\Sigma}}

\newcommand{\be}{\begin{equation}}
\newcommand{\ee}{\end{equation}}
\newcommand{\bea}{\begin{eqnarray}}
\newcommand{\eea}{\end{eqnarray}}
\newcommand{\bse}{\begin{subequations}}
\newcommand{\ese}{\end{subequations}}

\begin{document}
\title{Induced $p$-wave superfluidity in strongly interacting imbalanced Fermi gases}
\author{Kelly R. Patton}
\email[]{kpatton@lsu.edu}
\author{Daniel E. Sheehy}
\email[]{sheehy@phys.lsu.edu}
\affiliation{Department of Physics and Astronomy, Louisiana State University, Baton Rouge, Louisiana 70803 }
\date{November 19, 2010}
\begin{abstract} 
The induced interaction among the majority spin species, due to the presence of the minority species, is computed for
the case of a population-imbalanced resonantly-interacting Fermi gas. It is shown that this interaction leads to 
an instability, at low temperatures, of the recently observed polaron Fermi liquid phase of strongly imbalanced
Fermi gases to a $p$-wave superfluid state.  We find that the associated transition temperature, while quite small
in the weakly interacting BCS regime, is experimentally accessible in the strongly interacting unitary regime.
\end{abstract}
\maketitle 

The extraordinary controllability of cold atomic gases has
yielded a wide range of interesting phases of matter, including
a  bosonic Mott insulator and a paired superfluid state of two
species of atomic fermions~\cite{BlochRMP08,GiorginiRMP08}.    In the latter
setting, experiments have demonstrated control of both the interspecies
interactions and the relative density of the two spin states~\cite{Zwierlein2006,Partridge2006},
with the latter experimental knob being deleterious to pairing 
and superfluidity, which favors an equal density of the two
species.  

Thus, experiments on such imbalanced Fermi gases can probe
the stability of superfluidity in a correlated system and
therefore may shed light on the tendency towards pairing in 
related systems, such as electronic superconductors.  The phase
diagram of imbalanced Fermi gases is quite rich~\cite{ShinNature08,ShinPRL08}, possessing
regions of imbalanced superfluidity, phase separation, normal
Fermi liquid, and (possibly, though not yet observed) a region of
exotic Fulde-Ferrell-Larkin-Ovchinnikov (FFLO) superfluidity~\cite{Radzihovsky}. 

Our present focus is on the strongly imbalanced region, when the 
polarization $P = \frac{n_\uparrow - n_\downarrow}{n_\uparrow + n_\downarrow}$
(with $n_{\sigma}$ the density of fermion species $\sigma$)
is close to unity, with a small density of
spins-$\downarrow$ immersed in a spin-$\uparrow$ Fermi sea. 
Experiments in this regime~\cite{SchirotzekPRL09,Nascimbene} have found results consistent 
with the formation of spin polarons~\cite{ChevyPRA06}, in which a 
cloud of spins-$\uparrow$ gather around each spin-$\downarrow$, leading to a polaron
Fermi liquid state~\cite{CombescotPRL07,VeillettePRA08,CombescotPRL08}. 

The polaron theory of the strongly imbalanced Fermi gas predicts that
both the spins-$\uparrow$ and spins-$\downarrow$ are Fermi
liquids, exhibiting sharp Fermi surfaces at low 
temperature $T$. However, general arguments due
to Kohn and Luttinger (KL)~\cite{KohnPRL65,LuttingerPR66} predict
that such Fermi surfaces must be unstable to 
other ordered states as $T\to 0$.  A natural question then emerges: What is 
this ordered state for imbalanced Fermi gases?   Since the imposed imbalance 
(and concomitant Fermi-surface mismatch) precludes $s$-wave
singlet interspecies pairing, one instead expects
triplet (likely $p$-wave)  {\it intraspecies\/} pairing~\cite{KohnPRL65,LuttingerPR66,BulgacPRL06}
of the spin-$\uparrow$ and spin-$\downarrow$ fermions.

\begin{figure}
\vspace{-0.75cm}
\includegraphics[width=\columnwidth]{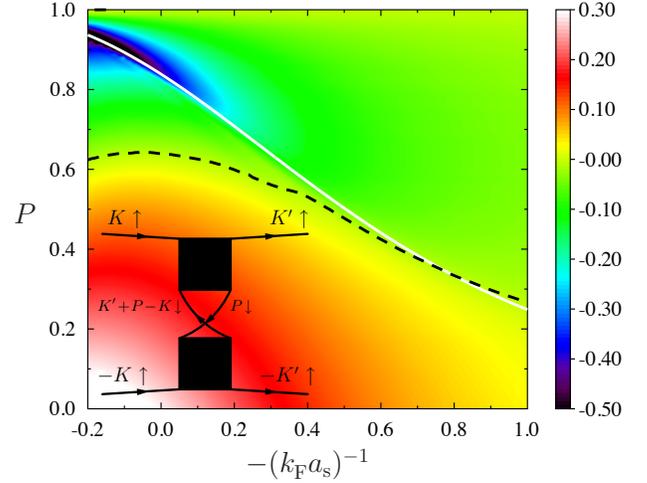}%
\vspace{-2.25cm}
\caption{(Color online)  The $p$-wave channel of the effective
interaction, determined by the diagrammatic inset, between majority spins, $v^{\ell=1}(\kfup,\kfup)$,
multiplied by the Fermi-energy density of states $N_{\uparrow}({\epsilon}^{}_{{\rm
F}\uparrow})$, as a function of the density
imbalance $P$ and $s$-wave scattering length $\as$, is shown.
Here, $k^{}_{\rm F}=(k^{}_{{\rm F}\uparrow}+k^{}_{{\rm
F}\downarrow})/2$. At large $P$, above the dashed line,
it is {\it attractive\/} leading to a $p$-wave superfluid at
temperatures below $T^{}_{\rm c}$.  The solid white line labels 
the location of a line of FFLO quantum critical points.  This coincides  with the location where $v^{\ell=1}(\kfup,\kfup)$ is large. 
 \label{fig1}} \vspace{-0.55cm}
\end{figure}

In this  Rapid Communication we develop a theoretical description of
the induced interactions among the majority spin-$\uparrow$ fermions
in the presence of a small density of spins-$\downarrow$, over a broad
range of interaction and polarization values.  We find 
an attractive effective interaction at the spin-$\uparrow$ Fermi surface,
shown in Fig.~\ref{fig1}, leading to an instability towards $p$-wave 
superfluidity below a transition temperature $\tc$, computed below~\cite{note1}.
In the extreme weak coupling Bardeen-Cooper-Schrieffer (BCS) limit,
where the $s$-wave scattering length $\as \to 0^{-}$, the $p$-wave
superfluid transition temperature  has been computed
\cite{BulgacPRL06}; unfortunately, it is exceptionally small (in
agreement with KL~\cite{KohnPRL65}), with $\tc\propto\exp[-c/(\kfup
\as)^2]$, where $\kfup$ is the spin-$\uparrow$ Fermi wave vector and
$c$ is a constant of order unity.  The pairing mechanism is quite
simple in this perturbative limit:\ density fluctuations of one
species leads to an attraction between particles of opposite spin.
For spins-$\uparrow$, this induced interaction is proportional to the
spin-$\downarrow$ density-density correlation function (i.e., the
Lindhard function).  Although a $p$-wave superfluid is predicted in the BCS limit, 
most experiments occur in the unitary
region where the interspecies interactions are strong, $|\as| \to
\infty$, invalidating simple perturbative results.  Our analysis of
induced interactions in  the unitary regime involves extending the
ladder approximation  (known to describe the polaron Fermi liquid
regime discussed above) to include subleading classes of Feynman
diagrams.  As seen in Fig.\ \ref{fig1}, the predicted effective
interaction in the $p$-wave channel  
can be quite large near the unitary regime 
%(and
%slightly into the Bose-Einstein condensate (BEC) side).  
We find a
maximum transition temperature of $k^{}_{\rm B}\tc \simeq 0.03
\epsilon^{}_{{\rm F}\uparrow}$, which is low but not unreasonable
given recently reported temperature scales  (e.g., Ref.~\onlinecite{Schirotzek2008}).

Our starting point is the standard one-channel model for two species ($\sigma = \uparrow,\downarrow$) of
fermion ($c_{\bk\sigma}^\dagger$) interacting via an $s$-wave Feshbach resonance~\cite{GurarieAnnalsPhys07}. The Hamiltonian is
\begin{equation}
\label{Hamiltonian} H=\sum_{{\bf k},\sigma}\xi^{}_{{\bf
k}\sigma}c^{\dagger}_{{\bf k}\sigma}c^{}_{{\bf
k}\sigma}+\frac{\lambda}{\sf V}\sum_{{\bf k},{\bf k}',{\bf
q}}c^{\dagger}_{{\bf k}+{\bf q}\uparrow}c^{\dagger}_{{\bf k}'-{\bf
q}\downarrow}c^{}_{{\bf k}'\downarrow}c^{}_{{\bf k}\uparrow},
\end{equation} where $\xi^{}_{{\bf k}\sigma}=\epsilon_{\bf
k}-\mu_{\sigma}$,  with dispersion $\epsilon_{\bf k} = k^2/2m$ ($\hbar=1$)  and chemical
potential $\mu_{\sigma}$; $\sf V$ is the system volume (henceforth set to unity), and $\lambda$
is the coupling strength  of a short-ranged 
pseudo-potential,
related to the experimentally controllable (via a magnetic field-tuned
Feshbach resonance) scattering 
length $\as$ via 
\begin{equation}
\label{renorm}
\frac{m}{4\pi \as}=\lambda^{-1}_{}+\sum_{\bf k}^{\Lambda}\frac{1}{2\epsilon_{\bf k}},
\end{equation} 
where $\Lambda$ is an ultraviolet cutoff; below we shall take the 
limit $\Lambda \to \infty$ while  $\lambda\rightarrow 0^{-}$, such that $\as$ 
remains fixed.

\begin{figure}
\includegraphics[width=0.85\columnwidth]{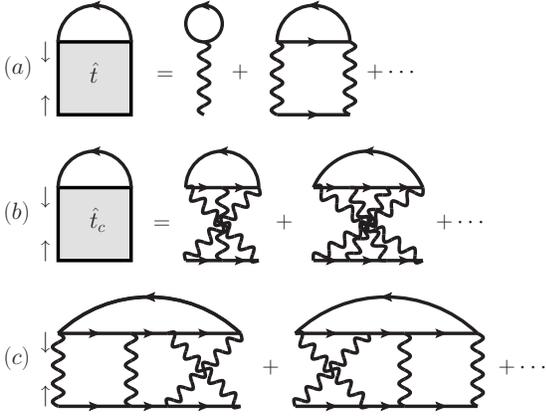}%
\caption{The ladder (a) and maximally crossed (b) diagrams contributing to the self-energy, which define
the ladder $\hat{t}$  and crossed-ladder $\hat{t}_c$ diagrams.  The bottom row, (c), shows a subleading class
of diagrams with both ladder and crossed-ladder diagrams that lead to a pairing instability.  The spin-$\uparrow$ lines are 
$2\times 2$ Nambu Green's functions $\hat{\curG}_{\uparrow}(K)$, spin-$\downarrow$ lines are  normal scalar Green's functions $G_\downarrow(K)$, and the wavy lines are $\lambda\sigma^{}_{z}$. 
}\label{fig2} \vspace{-0.35cm}
\end{figure}

We are interested in the phases of Eq.~(\ref{Hamiltonian}) in the strongly
imbalanced limit $P\to 1$ and proceed (in the spirit of mean-field theory) 
by assuming the presence of a self-consistently determined  
pairing amplitude $\Delta_\uparrow(\bk,\omega)$ among the spins-$\uparrow$, but not the spins-$\downarrow$~\cite{note1}. 
Under this assumption, the $2\times 2$ spin-$\uparrow$ Nambu Green's function, in Fourier-Matsubara space, is
$\hat{\curG}_\uparrow(K) =\begin{pmatrix} G_\uparrow(K) & F_\uparrow(K)\\
 F_\uparrow(K) &  -G_\uparrow(-K)\end{pmatrix} $, where the four-vector 
$K = (i\omega_{n},\bk)$.
$\hat{\curG}_\uparrow(K)$ satisfies the  Dyson equation~\cite{agd} $\hat{ \curG}^{-1}_\uparrow(K) =  \hat{\curG}_{\uparrow,0}^{-1}(K) -\Sigmah^{}_{\uparrow}(K)$, with 
bare Green's function $\hat{\curG}^{-1}_{\uparrow,0}(K) = \begin{pmatrix} i\omega_{n} - \xi_{\bk\uparrow} & 0 \\
0 &  i\omega_{n} + \xi_{\bk\uparrow}\end{pmatrix}$ and self-energy 
$\Sigmah_\uparrow(K)   =\begin{pmatrix} \Sigma_{\uparrow}(K) & -\Delta_\uparrow(K)\\
 -\Delta^*_\uparrow(K) &  -\Sigma_{\uparrow}(-K)\end{pmatrix} $.  Using the equation of 
motion~\cite{KadanoffBaym}, the self-energy can be expressed in terms of the two-particle vertex $\hat{\Gamma}$, by
\begin{widetext}
\bea
\Sigmah_\uparrow(K) = \lambda \sigma_z \sum_{K_1}  G_\downarrow(K_1)
\Big[ \sigma_0 -  \sum_{K_2} \hat{\curG}_{\uparrow}(K_2)G_\downarrow(K+K_1-K_2) \hat{\Gamma} (K_2,K+K_1-K_2,K_1,K)\Big],
\label{eq:fullself}
\eea
\end{widetext}
 where $\sigma_0$ is the $2\times 2$ identity matrix, $\sigma_z$ is a Pauli matrix, and  $G_\downarrow(K)$
is the scalar spin-$\downarrow$ Green's function, which satisfies a similar expression,
but with a diagonal self-energy (since we assumed the spins-$\downarrow$ are unpaired). The summation 
 is short for $\sum_K \equiv k^{}_{\rm B}T \sum_{i\omega_{n}}\sum_\bk$.

Although Eq.~(\ref{eq:fullself}) is in principle
exact, to make progress we must make a physically motivated approximation for $\hat{\Gamma}$, correpsonding
to certain classes of Feynman diagrams.  Previous work has analyzed the phases of imbalanced
Fermi gases within the $t$-matrix or ladder approximation~\cite{CombescotPRL07,Punk,Schneider}.
The set of diagrams associated with the ladder approximation also emerges
 in the large-$N$ approximation~\cite{VeillettePRA08}, in which one generalizes the model to consist of $2N$ species of
fermion. 
Within the present formalism in which $\hat{\curG}_\uparrow(K)$ has Nambu structure, these
contributions arise from including the ladder plus the maximally crossed self-energy diagrams, sketched 
in Figs.~\ref{fig2}(a) and \ref{fig2}(b), respectively.  We first analyze Eq.~(\ref{eq:fullself}) including only these
diagrams.   Exchanging $\lambda$ for $\as$ using Eq.~(\ref{renorm}), 
and keeping only contributions that are finite in the limit $\Lambda \to \infty$, 
we find a {\it diagonal\/} self-energy;
\be
\Sigmah_\uparrow(K) = \sum_{Q}  G_\downarrow(Q)
\begin{pmatrix}
t(Q+K) & 0 \\
0 & -t(Q-K)
\end{pmatrix} ,
\label{Eq:normalsigma}
\ee
indicating the absence, at this level, of pairing for the spins-$\uparrow$.  Here,   
\be
\label{eq:tee1}
t(K)^{-1} =\frac{m}{4\pi \as} + \sum_Q G_\downarrow(K-Q) G_\uparrow(Q)-\sum_{{\bf q}}\frac{1}{2\epsilon_{\bf q}},
\ee
is the usual scalar $t$-matrix.
%\begin{figure}
%\includegraphics[scale=0.75]{fig3.eps}%
%\caption{Feynman diagram for  Eq.~(\ref{eq:effectiveinteraction}), the effective interaction among the majority 
%spin-$\uparrow$ fermions mediated by the spin-$\downarrow$ fermions.
%The $t$-matrices are given by Eq.~(\ref{eq:tee1}).
% }\label{figv}\vspace{-0.5cm}
%\end{figure} 

Thus, the contributions from Figs.~\ref{fig2}(a) and \ref{fig2}(b) yield an unpaired solution for the self-energy, i.e., a Fermi liquid,
as found by  
previous $t$-matrix or 
large-$N$ theories~\cite{CombescotPRL07,Punk,VeillettePRA08,Schneider}.  Given
that the KL arguments imply the eventual instability of this state, we now turn to subleading 
contributions to the self-energy, shown in Fig.~\ref{fig2}(c),  
that possess ladder and crossed-ladder subdiagrams.  Again replacing
 $\lambda$ for $\as$ using Eq.~(\ref{renorm}), we find that these diagrams yield an off-diagonal contribution to the self-energy, i.e.,\ a pairing amplitude $\Delta_\uparrow(K)$ 
given by 
\bea
 &&\Delta_\uparrow(K)= \sum_{Q}V(K,K')   F_\uparrow(K'),
\label{eq:deltaself}
\\
&&\hspace{-.65cm}V\!(K,K')\!=\!\!\sum_{P}  t(P\!-\!K) t(\!P\!+\!K') G_\downarrow(\!P\!)G_\downarrow(K'\!+\!P-\!K\!).
\label{eq:effectiveinteraction}
\eea
Here, $V(K,K')$  is the  effective induced interaction among spins-$\uparrow$; the corresponding Feynman 
diagram is shown in Fig.~\ref{fig1}. In principle, the integral equation (\ref{eq:deltaself}) must be solved self-consistently along with the  
diagonal self-energy, Eq.\ (\ref{Eq:normalsigma}).  However, we shall we use some physically motivated
approximations to simplify our analysis, focusing on the onset of pairing of the spins-$\uparrow$ 
at a temperature $\tc$ (above which the system is a Fermi liquid).
 We assume the presence of a static momentum-dependent pairing order parameter,
$\Delta^{}_{\uparrow}(K) = \Delta^{}_{\uparrow}(\bk)$, and neglect the frequency dependence of $V(K,K')$~\cite{LuttingerPR66}.
For the diagonal components of the self-energy, we simply assume that the chemical potential is
renormalized to the Fermi energy via $\mu^{}_{\sigma}\rightarrow\mu_{\sigma}-\Sigma_{\sigma}({\bf
k}^{\sigma}_{\rm F},0)=\epsilon_{{\rm F}\sigma}$ (consistent with the Luttinger theorem~\cite{LuttingerWard,SachdevYang}). 
Within these approximations and after analytic continuation the effective interaction takes the form 
\begin{align}
\label{effectiveinteraction}
V(\bk,\bk')\approx 2{\rm Re}&\big[\sum_{{\bf q}}t^{\rm r}_{}(\bk+{\bf q},\xi_{{\bf q}\downarrow}){t}^{\rm a}_{}({\bf q}-\bk',\xi_{{\bf q}\downarrow})
\\& \times G^{\rm r}_{\downarrow}(\bk-\bk'+{\bf q},\xi_{{\bf q}\downarrow}) 
n^{}_{\rm F}(\xi_{{\bf q}\downarrow})\big],\nonumber
\end{align}
 where r/a refers to the retarded or advanced quantities
and $n^{}_{\rm F}(\omega)$ is the Fermi distribution function.  Equation (\ref{eq:deltaself}) then simplifies to
\be
\label{Eq:gaptc}
\Delta_\uparrow(\bk) = -\sum_{\bk'}V(\bk,\bk') \frac{\Delta_{\uparrow}(\bk')}{2E_{\bk'}}\tanh\frac{E_{\bk'}}{2T},
\ee
with $E_\bk = \sqrt{\xi_{\bk\uparrow}^2+|\Delta_\uparrow(\bk)|^2}$, 
the solution of which requires an understanding of the momentum structure of the effective interaction $V(\bk,\bk')$ in the vicinity of 
the spin-$\uparrow$ Fermi surface.  
The transition temperature $\tc$ for a given angular momentum is found by solving the linearized, in $\Delta^{}_{\uparrow}(\bk)$, version of Eq.~(\ref{Eq:gaptc}) and projecting onto 
the
relevant channel \cite{GurarieAnnalsPhys07}. Assuming $p$-wave pairing, one needs the $\ell = 1$ projection of
the induced interaction,  $v^{\ell=1}(k,k')=\int_{0}^{\pi}d\theta\,\sin\theta\cos\theta V(\bk,\bk')$, 
where $\theta$ is the angle 
between $\bk$ and $\bk'$.  Furthermore, we find via a direct numerical integration of Eq.~(\ref{effectiveinteraction}), that 
 $v^{1}(k,k')$ is only appreciable for $k$ and $k'$ within a range $~\kfdown$ of each other; this defines an effective bandwidth, 
of the order of $\epsilon_{{\rm F}\downarrow}$, over which the induced interaction is nonzero.

\begin{figure}
\includegraphics[width=\columnwidth]{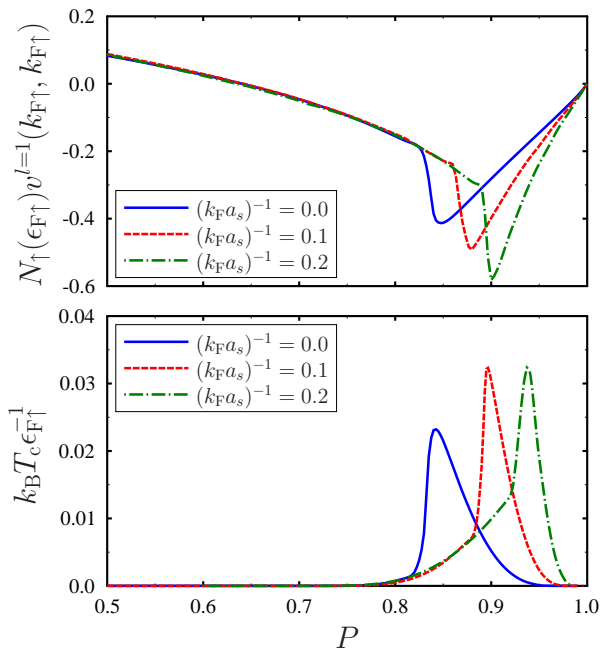}%
\vspace{-0.5cm}
\caption{The top panel shows the $l=1$ channel of the effective interaction for spin-$\uparrow$ fermions, times the density of states, as a function of polarization, at unitary and into the BEC side. The bottom panel shows the corresponding $p$-wave transition temperature,
according to Eq.~(\ref{eq:tcresult}), with  $\efup/\efdown = [(1+P)/(1-P)]^{2/3}$, via a numerical integration of Eq.~(\ref{effectiveinteraction}).  
 }\label{fig3} \vspace{-0.5cm}
\end{figure}

The strong induced attraction among the spins-$\uparrow$, shown in  Figs.~\ref{fig1} and \ref{fig3}a, suggest
a robust $p$-wave superfluid at $T\to 0$; to estimate the associated transition temperature $\tc$ we must
make further approximations.
  The remaining momentum integrations in Eq.~(\ref{Eq:gaptc}) are sharply peaked at the Fermi surface, yielding the result 
\be
\label{eq:tcresult}
k^{}_{\rm B}\tc \approx \frac{2{\rm e}^{\gamma}}{\pi}\epsilon_{{\rm F}\downarrow} \exp\Big[ \frac{1}{N_{\uparrow}(\efup) v^{\ell=1}(\kfup ,\kfup) }\Big] ,
\ee
with $\gamma$ the Euler gamma constant. 
 We have also found~\cite{unpublished} the same result for the transition temperature  
(and the same effective interaction, 
Eq.~(\ref{eq:effectiveinteraction})) via a somewhat different approach by considering the Thouless criterion~\cite{Thouless} for 
$\tc$, determined by the point at which the spin-$\uparrow$ pair-pair fluctuations in the normal state become unbounded.  Within 
such an approach, Eq.~(\ref{eq:effectiveinteraction}) is the irreducible vertex in the particle-particle channel of the Bethe-Salpeter 
equation~\cite{Nozieresbook}.

In general, \eqref{effectiveinteraction} has to be determined
numerically, but analytic results can be found for certain limiting
cases.   In the asymptotic BCS limit 
$\as \rightarrow 0^{-}$,  the $t$-matrix $t^{{\rm r}/{\rm a}}(\bk,\omega) \rightarrow 4\pi \as/m$,
and Eq.~(\ref{effectiveinteraction}) reduces to the result of Ref.\ \cite{BulgacPRL06}
(where, as we have noted, $\tc$ is extremely small).  Analytic results can also
be obtained in the  extremely imbalanced
limit, i.e.,\ $\kfup/\kfdown \gg 1$. In this limit the $t$-matrices appearing in Eq.~(\ref{effectiveinteraction}) also become independent of ${\bf k}$ and ${\bf k}'$;  $t^{{\rm r}/{\rm a}}(\kfup \pm{\bf
q},\xi_{{\bf q}\downarrow})\rightarrow\big(\frac{m}{4\pi \as}-\frac{m \kfup}{4\pi^{2}}\big)^{-1}$. 
% The remaining integral again leads to the Lindhard function, giving
Evaluating the remaining integral gives
\begin{equation}
k^{}_{\rm B}\tc\approx \frac{2{\rm e}^{\gamma}}{\pi}\epsilon_{{\rm F}\downarrow}
\exp\Big[-\frac{3}{2}\frac{z}{\ln(z)}\Big(\frac{\pi}{2k^{}_{{\rm F}\downarrow}\as}-\frac{z}{2}\Big)^{2}\Big],
\end{equation} with $z=\kfup/\kfdown$.   This formula correctly
captures the vanishing of $\tc$ for $P\to 1$, but doesn't adequately
capture the peaks shown in  Fig.~\ref{fig3}, which where found by a
direct numerical analysis of Eq.~(\ref{effectiveinteraction}). These
results indicate the presence of pairing at an experimentally
accessible temperature in unitary imbalanced gases. 

%The peak in $T_{\rm c}$, as a function of $P$, is present even in the
%perturbative regime \cite{BulgacPRL06}.  At high polarization, the
%particle-hole excitations responsible for the induced interaction
%\eqref{effectiveinteraction}, with a transferred momentum larger than
%the diameter of the spin-$\downarrow$ Fermi surface are energetically
%suppressed, while at lower imbalance the particle-hole spectrum
%becomes approximately flat, leading to a vanishingly small $p$-wave
%projection. 
%

The peak in the induced attraction, and in $T_{\rm c}$, at large $P$,
can be understood by noting that a crucial contribution comes from
particle-hole excitations at the spin-$\downarrow$ Fermi surface.
However, particle-hole excitations with a transferred momentum
larger than $2\kfdown$ are energetically
suppressed, so that the associated density response exhibits a strong
momentum dependence at $k\simeq 2\kfdown$, and a correspondingly
large $p$-wave projected interaction at $\kfup \simeq 2\kfdown$, or
$P = \frac{\kfup^3-\kfdown^3}{\kfup^3+\kfdown^3} \simeq 0.78$.
 Near unitarity, the magnitude of this peak is enhanced by
the presence of strong fluctuations toward the FFLO phase, signaled by
a divergence of the  $t$-matrix at a non-zero  wave vector ${\bf Q}$,
i.e., $t^{\rm r}({\bf Q},0)\rightarrow\infty$. 
To illustrate this, in 
Fig.~\ref{fig1} the white line shows the $P$ at which a quantum phase transition to the FFLO state occurs;
thus, below this line, the $p$-wave paired phase may undergo a second phase transition to the FFLO.

The confirmation of our scenario will require detecting the onset of $p$-wave pairing at $\tc$ and
the properties of the resulting $p$-wave superfluid below $\tc$. 
This can be done via standard probes
of superfluidity, such as the presence of vortices in a rotating cloud~\cite{Zwierlein2005}.
 Following general 
arguments~\cite{AndersonMorelPR61,GurariePRL05,NishidaAnalPhys09}, we expect a $p_x+ip_y$ ground state, i.e.\  $\Delta_\uparrow({\bf k})=\Delta_{0}Y_{1,1}(\hat{\bf k})$.
The anisotropic gapping of the spin-$\uparrow$ Fermi surface should yield a signature 
in radio-frequency (RF) spectroscopy, which measures the atom transfer rate of one spin species from the
interacting system to an unoccupied energy level \cite{ChenRepProgPhys09}, probing the 
spectral function.  However, we find that the associated peak position in the RF line-shape is at $\omega \simeq \Delta_{0} (\Delta_0/\efup)$, 
a very small energy scale given the smallness of $\tc$ computed above.  A more promising route,
that we leave for future research, is the question of how the onset of pairing impacts the formation of 
the spin-$\downarrow$ polarons (as reflected in, e.g., the spin-$\downarrow$ RF spectra~\cite{SchirotzekPRL09}). 

In summary,  we have calculated the induced interaction between like
atoms in the normal state of an imbalanced two-component Fermi gas.
In the absence of any competing instabilities (which certainly occur at smaller $P$,
where the regimes of magnetic superfluidity~\cite{SheehyAnalPhys07}, phase separation and,
possibly FFLO phase occur), this interaction leads
to the formation of a $p$-wave superfluid in the majority spin species, with a transition temperature
that peaks, for $P$ close to unity, at a few percent of the spin-$\uparrow$ Fermi energy. 

This work has been supported by the Louisiana Board of Regents, under grant LEQSF (2008-11)-RD-A-10.


\begin{thebibliography}{99}
\bibitem{note1} Within our theory, the spins-$\uparrow$ are also predicted to induce pairing among
the spins-$\downarrow$; we find the associated temperature scale to be much lower.
\end{thebibliography}
\end{document}